\documentclass[print,12pt]{elsarticle}

\biboptions{sort&compress}

\usepackage{graphicx}
\usepackage{amssymb}

\usepackage{lineno}
\usepackage{ulem}

\journal{Nucl. Phys. A}

\begin{document}

\begin{frontmatter}


\title{Role of charged particle emission on the evaporation residue formation in the $^{82}$Se+$^{138}$Ba reaction leading to the $^{220}$Th compound nucleus}



\author[chib,ct]{G.~Mandaglio\corref{mycorrespondingauthor}}
\cortext[mycorrespondingauthor]{Corresponding author}
\ead{gmandaglio@unime.it}
\author[dub,uzb]{A. K. Nasirov\corref{mycorrespondingauthor}}
\cortext[mycorrespondingauthor]{}
\ead{nasirov@jinr.ru}
\author[mift]{ A. Anastasi} 
\author[lnf]{F. Curciarello}
\author[mift]{ G. Fazio}
\author[mift]{G. Giardina}

\address[chib]{Dipartimento di Scienze Chimiche, Biologiche, Farmaceutiche ed Ambientali, University of Messina, Messina, Italy}
\address[ct]{INFN Sezione di Catania, Catania, Italy}
\address[dub]{JINR - Bogoliubov Laboratory of Theoretical Physics, Dubna, Russia}
\address[uzb]{Institute of Nuclear Physics, Academy of Science of Uzbekistan, Tashkent, Uzbekistan}
\address[mift]{Dipartimento di Scienze Matematiche e Informatiche, Scienze Fisiche e Scienze della Terra, University of Messina, Messina, Italy}
\address[lnf]{INFN Laboratori Nazionali di Frascati, Frascati, Italy}


\begin{abstract}
We present detailed results of a theoretical investigation on the production of evaporation residue
 nuclei obtained  in a heavy ion reaction when charged particles (proton and
 $\alpha$-particle) are also emitted  
  with the neutron evaporation along the deexcitation 
 cascade of the formed compound nucleus.  
 The  almost mass symmetric $^{82}$Se+$^{138}$Ba reaction  has been studied 
 since  there are  many experimental results on individual evaporation 
 residue (ER) cross sections after few light particle emissions along the cascade of
the $^{220}$Th compound nucleus (CN) covering the wide 12--70 MeV excitation 
energy range. 
  Our specific theoretical results on the ER cross sections for the $^{82}$Se+$^{138}$Ba are in good
   agreement with the available experimental measurements, but our overall
    theoretical results  concerning all possible relevant contributions of evaporation residues are several times greater than  the ERs measured in experiment. The discrepancy could be due to the experimental difficulties in the identification of ER nuclei after the emission of multiple neutral and charged particles, nevertheless the analysis of ER data is very important to test the reliability of the model and to stress the importance on the investigation of ER nuclei also obtained after charged particle emissions.
\end{abstract}


\end{frontmatter}



\section{Introduction}
The study of nuclear reactions continues to be of great interest in the scientific community to better understand the  mechanism of  the formation of final products 
  in a nuclear collision.  There are still  relevant unclear discrepancies between experimental results 
 as well as between different theoretical procedures \cite{NPA17}. 
  Compound nucleus (CN) is formed if dinuclear system (DNS)  \cite{volk} survives against quasifission which is 
  dominant process in almost mass symmetric reactions.
  Compound nucleus stage can not be reached  for angular momentum values $\ell>\ell_{\rm max}$ (where $\ell_{\rm max}$ is the maximum value of the angular momentum contributing to the DNS formation \cite{NPA17})
and   the fast fission occurs producing binary fission-like fragments.
At each step along the deexcitation cascade of the excited compound nucleus (CN) by emission of  light particles in competition with the fission process, the evaporation residue (ER) nuclei  
can be formed \cite{NPA17,PRC91,FazioJP77,aglio2009,aglio2012} as  reaction products. 
In this complex context two aspects of experimental uncertainties can be stressed: 
i) quasifission, fast fission and fusion-fission products might 
be overlapped; ii) some ER nuclei can not be detected and identified due to an unavoidable limits of 
experimental setup causing difficulties in estimations
of the cross sections by analysis of data \cite{PRC652002,JPCS18}. 
In fact, in the case of the $^{82}$Se+$^{138}$Ba reaction \cite{PRC652002} the individual experimental ER contributions are in general well distinguishable, in some cases the ER channels are detectable as sum of two indistinguishable  contributions; in other cases, other  ER contributions that are relevant according to our estimations
 have not been measured.  

In the analysis of experimental data there are unavoidable uncertainty  on the identification 
and separation of the products that are formed in  different channels
 of the reaction. Of course, 
  even the theoretical models   are not free from 
  serious uncertainties of the obtained results due to the assumptions made 
  in their formulation.

 In this paper, we present the results of calculation of the individual ER excitation 
 functions for the $^{82}$Se+$^{138}$Ba almost mass symmetric reaction (characterized by  a very  low mass asymmetry parameter value $\eta=0.255$) since it is possible to explore a wide region  of excitation energy $E^*_{\rm CN}$
  from  12 MeV (corresponding to the $E^*_{\rm thr}$ threshold energy for this entrance channel leading to the  $^{220}$Th CN formation) up to 70 MeV
of excitation energy of CN, when the emission of the charged particles (proton and $\alpha$) are also considered together with emission of neutrons.
Therefore, the study of the $^{82}$Se+$^{138}$Ba reaction remains a very 
 useful opportunity to analyze  the ER formation from lower excitation energies of CN.
 Moreover, the theoretical study of the $^{82}$Se+$^{138}$Ba reaction benefits of the large set of experimental data \cite{PRC652002} available for the  individual excitation function of evaporation residue   which can be compared with our 
 theoretical results and discussed.

This large set of experimental data is a good opportunity to look for the necessary improvemnts in the experimental and theoretical investigations on the formation of ER nuclei also taking into account the   various combinations of charged particle  emissions ($\alpha$ and proton). In this context, it is also possible to obtain useful information on the ratio between the total evaporation residue cross section (charged and neutral particle emissions) and the one produced by neutron emissions only \cite{arxivprc} at different values of $E^*_{CN}$.
    
     In Section II we present the main procedures and the necessary formulas to calculate  the observable cross sections by the non  adiabatic approach.
In Section III the calculated results of the individual cross sections of ER nuclei 
for the $^{82}$Se+$^{138}$Ba reaction are compared with the experimental ones.
  In Section IV we give our conclusions.

\section{Theoretical procedures}

The reaction mechanism of heavy ion collisions near the Coulomb barrier energies 
is considered  as the formation of a dinuclear system at the nuclear contact of reactants with a continuous exchange of nucleons between the constituent nuclei of DNS during its lifetime; then, this DNS can evolves into a complete fusion of its constituent nuclei in competition with the possibility of a separation of its components (quasifission process). In the former case, the complete fusion system can evolves into the statistically equilibrated CN  formation, in competition with the fast fission process (for angular momentum values of CN $\ell>\ell_{\rm cr}$ at which the fission barrier $B_{\rm fis}$ is equal to zero) that leads to formation of two fragments.
Therefore, only the deexcitation of CN and other intermediate excited nuclei reached along the deexcitation cascade after emission of light particles (n, p, $\alpha$-particle, and $\gamma$-quanta) being survived fission can form the evaporation residue nuclei in competition with the fission process for each excited nucleus.

The cross sections of the related reaction mechanisms are calculated by the following relations:

\begin{eqnarray}
\sigma_{\rm cap}(E_{\rm c.m.};\alpha_1,\alpha_2) & = &\sum^{\ell_d(E_{\rm c.m.})}_{\ell=0} \sigma_{\rm cap}^{\ell}(E_{\rm c.m.},\ell;\alpha_1,\alpha_2)\nonumber \\
& =& \frac{\lambda^2}{4\pi}\sum^{\ell_d(E_{\rm c.m.})}_{\ell=0}(2\ell+1)\nonumber\\
& \times&\mathcal{P}^{\ell}_{\rm cap}(E_{\rm c.m.},\ell;\alpha_1,\alpha_2),
\label{eqcapsum}
\end{eqnarray}
where the capture cross section $\sigma_{\rm cap}$  is determined by the number
of partial waves which lead to the path  of
colliding nuclei to be trapped in the well of the nucleus-nucleus
potential. The size of the potential well decreases with
increasing orbital angular momentum $\ell$.
The partial capture cross section $\sigma_{\rm cap}^{\ell}$ is the sum of the partial complete fusion  $\sigma_{\rm fus}(E_{\rm c.m.};\alpha_1,\alpha_2)$ and quasifission $\sigma_{\rm qf}(E_{\rm c.m.};\alpha_1,\alpha_2)$ cross sections.
In this formula $\mathcal{P}^{\ell}_{\rm cap}$ is the capture probability which
depends on the collision dynamics and  is 1 for $\ell_{\rm min}\leq\ell\leq\ell_d$, while is 0 if $\ell<\ell_{\rm min}$ or $\ell>\ell_d$ because the friction coefficient is not so strong to trap the projectile in the potential well; $\alpha_1$ and $\alpha_2$ are the angles of   the symmetry axes of deformed colliding nuclei relative to the direction of motion\cite{PRC91,NPA17}.
Moreover, the maximal values of partial waves $\ell_{\rm d}$ (leading to capture) is calculated by the solution of the equation of the relative motion of nuclei \cite{PRC72,NuclPhys05,FazioJP72}, and $\ell_{\rm min}$ is the minimal value of $\ell$ leading to capture.

The complete fusion (CF) cross section of the deformed mononucleus is obtained as:
\begin{eqnarray}
\sigma_{\rm CF}(E_{\rm c.m.};\alpha_1,\alpha_2)&=&\sum^{\ell_d(E_{\rm c.m.})}_{\ell=0} \sigma_{\rm cap}^{\ell}(E_{\rm c.m.},\ell;\alpha_1,\alpha_2)\nonumber\\ &\times& P_{CF}^{\ell}(E_{\rm c.m.},\ell;\alpha_1,\alpha_2),
\label{eqfus}
\end{eqnarray}
while the quasifission cross section
 is obtained as the complementary part of (2)
\begin{eqnarray}
\sigma_{\rm qf}(E_{\rm c.m.};\alpha_1,\alpha_2)&=&\sum^{\ell_d(E_{\rm c.m.})}_{\ell=0} \sigma_{\rm cap}^{\ell}(E_{\rm c.m.},\ell;\alpha_1,\alpha_2)\nonumber \\ &\times& [1-P_{CF}^{\ell}(E_{\rm c.m.},\ell;\alpha_1,\alpha_2)].
\label{eqqfis}
\end{eqnarray}

The competition between complete fusion and quasifission processes during the DNS evolution is determined by the complete fusion probability $P_{CF}^{\ell}$ which is calculated by the expression \cite{NuclPhys05,FazioJP72}
\begin{eqnarray}
\label{pcne} P_{\rm CF}^\ell(E^*_{\rm DNS},\ell; \{\alpha_i\})&=&\sum_{Z_{\rm sym}}^{Z_{\rm max}}
P^{(Z)}_{\rm CF}(E^*_{\rm DNS},\ell; \{\alpha_i\})\nonumber \\
 Y_Z(E^*_{\rm DNS},\ell),
\end{eqnarray}
where $Z_{sym}$=$(Z_1+Z_2)/2$ and $Z_{max}$ corresponds to the point where the driving potential reach its maximum, i.e., the value to which the intrinsic fusion barrier $B^*_{\rm fus}=0$, $Y_Z(E^*_{\rm DNS},\ell)$ is the charge distribution function (see Appendix of Ref. \cite{prc84}) operating on the $P_{\rm CF}^{(Z)}$ factor. The mass and charge distribution among the DNS fragments are calculated by solving the transport master equation \cite{epja34,jolos}. Equation (\ref{pcne}) allows us to take into account the fusion probabilities from the DNS charge asymmetry configurations which differ from the charge numbers of the projectile and target nuclei. 
The DNS lifetime with the given charge asymmetry $Z=Z_1$ and $Z_2=Z_{\rm CN}-Z$ depends on the depth  of the potential well which is quasifission barrier $B_{\rm qf}^{(Z)}$ and its excitation energy $E^{(Z)*}_{\rm DNS}$. Therefore, at the change of the charge asymmetry by the nucleon transfer in
 DNS, its evolution is influenced by the competition between the reaction mechanism of separation of the nuclei constituting the DNS (quasifission process) and the exchange process of various nucleons tending to reach the complete fusion of the nuclei in the DNS. A decisive role for this competition   is played by the intrinsic fusion $B^*_{\rm fus}$ and  the quasifission barrier $B_{\rm qf}$  whose values are determined by the charge asymmetry and angular momentum $\ell$ of DNS. Therefore, we take into account the change of charge asymmetry by nucleon transfer before the decay of DNS \cite{epja13}. It has been observed \cite{NPA17,aglio2012,PRC72,prc79,prc84} that as the angular momentum increases, the quasifission barrier $B_{\rm qf}$ decreases. On the other hand, the quasifission barrier $B_{\rm qf}$ also decreases with the decrease of DNS charge asymmetry due to the increase of the Coulomb interaction. Thus the stability of the DNS against its decay in two nuclei (quasifission process) decreases to the decrease of the $B_{\rm qf}$ barrier, and consequently also the probability of the complete fusion $P_{\rm CF}$  decreases~\cite{PRC91,PLB2010}. 
Therefore, the part of the complete fusion cross section $\sigma_{\rm CF}$ (of the deformed mononucleus) that is transformed into the  compound nucleus cross section $\sigma_{\rm CN}$ (of the statistically equilibrated system CN)  is obtained as
\begin{eqnarray}
\sigma_{\rm CN}(E_{\rm c.m.};\alpha_1,\alpha_2)&=&\sum^{\ell_{\rm cr}}_{\ell=0} \sigma_{\rm cap}^{\ell}(E_{\rm c.m.},\ell;\alpha_1,\alpha_2)\nonumber\\ &\times& P_{CF}^{\ell}(E_{\rm c.m.},\ell;\alpha_1,\alpha_2),
\label{eqfus}
\end{eqnarray}
while the part going in fast fission is related to the angular momentum interval from $\ell_{\rm cr}$ to $\ell_{\rm d}$
\begin{eqnarray}
\sigma_{\rm ff}(E_{\rm c.m.};\alpha_1,\alpha_2)&=&\sum^{\ell_{\rm d}(E_{\rm c.m.})}_{\ell=\ell_{\rm cr}} \sigma_{\rm cap}^{\ell}(E_{\rm c.m.},\ell;\alpha_1,\alpha_2)\nonumber\\ &\times& P_{CF}^{\ell}(E_{\rm c.m.},\ell;\alpha_1,\alpha_2);
\label{eqff}  
\end{eqnarray}
therefore, the compound nucleus probability $P_{\rm CN}(E_{\rm CN}^*)$ corresponds to the ratio  $\sigma_{\rm CN}/\sigma_{\rm cap}$ between the cross section of the CN formation and the one of the capture process.

 The partial cross sections of the CN formation are used to calculate evaporation residue cross sections  at given values of
the excitation energy $E_{\rm CN}^*$ and angular momentum $\ell$,
 for each successive intermediate excited nucleus, formed at $x$th step along the
deexcitation cascade with excitation energy $E_{x}^*$,  by the advanced statistical model \cite{aglio2012},

\begin{equation}
\sigma_{\rm ER}^{(x)}(E^*_{x})=\Sigma^{\ell_d}_{\ell=0}
(2\ell+1)\sigma_{\rm ER}^{(x)}(E^*_{x},\ell),
\label{equER}
\end{equation}
where $\sigma_{\rm ER}^{(x)}(E^*_{x},\ell)$ is the partial cross section of ER
formation obtained  after the emission
of particles $\nu(x)$n + $y(x)$p + $k(x)\alpha$ + $s(x)$
(where $\nu(x)$, $y$, $k$, and $s$ are numbers of neutrons, protons,
$\alpha$-particles, and $\gamma$-quanta, respectively)  from the  intermediate
nucleus with excitation energy $E^*_{x}$   at each step $x$
of the deexcitation cascade by the formula
(for more details, see papers \cite{FazioMPL2005,aglio2012,PRC72}):
\begin{equation}
\label{SigmaEN}
\sigma_{\rm ER}^{(x)}(E^*_{x},\ell)=
\sigma^{(x-1)}_{\rm ER}(E^*_{x-1},\ell)W^{(x)}_{\rm sur}(E^*_{x-1},\ell).
\end{equation}

In equation (\ref{SigmaEN}),  $\sigma_{\rm ER}^{(x-1)}(E^*_{x-1},\ell)$
 is the partial cross section
of the intermediate excited nucleus formation at the $(x-1)$th step, and
$W^{(x)}_{\rm sur}$ is the survival probability of the
$x$th intermediate nucleus against fission along each step of the
deexcitation cascade of  CN.
In calculation of the $W_{\rm sur}^{(x)}(E^*_{x-1},\ell)$ the used fission barrier is
 the sum of the parametrized macroscopic fission
barrier $B_{fis}^{m}$ \cite{SierkPRC33}
 and  the microscopic  correction $\delta W = \delta W_{sad} -\delta W_{gs}$ due to shell effects;
 by considering the large deformation of a fissioning nucleus at the saddle point, $\delta W_{sad}$ is much smaller than the $\delta W_{gs}$ value and the  $\delta W$
 can be expressed as
 $\delta W \cong -\delta W_{\rm gs} $.

 Therefore, the effective fission barrier, as a function of $\ell$ and $T$ for each excited nucleus formed at various steps along the deexcitation cascade of CN, is calculated by the expression
\begin{equation}
\label{fissb} B_{\rm fis}(\ell,T)=    B_{fis}^{m}-h(T) \ q(\ell) \ \delta
W,
\end{equation}
where $ B_{fis}^{m}$ is the macroscopic term \cite{SierkPRC33}, and
$h(T)$ and $q(\ell$) represent the damping functions of the nuclear
shell correction with the increase of the
excitation energy $E^{*}$  and $\ell$ angular momentum, respectively \cite{aglio2012}:
\begin{equation}
h(T) = \{ 1 + \exp [(T-T_{0})/ d]\}^{-1}
\label{hoft}
\end{equation}
and
\begin{equation}
q(\ell) = \{ 1 + \exp [(\ell-\ell_{1/2})/\Delta \ell]\}^{-1}.
\label{hofl}
\end{equation}
In Eq. (\ref{hoft}), $T=\sqrt{E^*/a}$ represents the nuclear
temperature depending on the excitation energy
$E^*$ and the level density parameter $a$, $d= 0.3$~MeV is the rate of
washing out the shell corrections with
the temperature, and $T_0=1.16$~MeV is the value at which the damping
factor $h(T)$
is reduced by 1/2. Analogously, in Eq. (\ref{hofl}), $\Delta \ell
=3\hbar$ is the rate of washing out the shell corrections with the angular momentum, and $\ell_{1/2}
=20\hbar$ is the value
at which the damping factor $q(\ell)$ is reduced by 1/2.

In this context, for the intrinsic level density parameter $a$ we use the general expression \cite{Ignatyuk}
especially tailored to account for the shell effects in the level density

\begin{equation}
a(E^*)= \tilde{a}\left\{ 1+\delta W \left[\frac{1-exp(-\gamma E^*)}{E^*}\right]     \right\}
\label{level_density_a}
\end{equation}
where $\tilde a = 0.094 \times A$  MeV$^{-1}$ is the asymptotic value that takes into 
account the dependence on the mass number $A$,
and $\gamma=$0.0064 MeV$^{-1}$ is the parameter which accounts for the rate at which shell effects wash out with
excitation energy for neutron or other light particle emission.

Moreover, in order to determine the $a_{\rm fis}$ level density parameter in the
fission channel we use the relation $a_{\rm fis}(E^*) =a_n(E^*)\times r(E^*)$ found in \cite{darrigo94} where $r(E^{*})$ is given by the relation

\begin{equation}
r(E^*)=\frac{\left[exp(-\gamma_{fis} E^*) - \left(1+\frac{E^*}{\delta W}\right) \right]}{\left[exp(-\gamma E^*) - \left(1+\frac{E^*}{\delta W}\right) \right]}
\label{ratio_a_parameters}
\end{equation}
with $\gamma_{fis}=0.024$~MeV$^{-1}$.

We point out that relation (\ref{ratio_a_parameters}) allows one to describe in a consistent approach including
collective effects the important function $a_{\rm fis}(E^{*})/a_{\rm n}(E^{*})$ ratio given by a general expression $r(E^{*})$,
rather than adjust by a phenomenological way the value of the cited $a_{\rm fis}/a_{\rm n}$ ratio for each excited nucleus.
Therefore, this procedure allows the shell corrections to become sensitive to the excitation energy $E^*$, while for the intrinsic level density $\rho_{\rm int}(E^{*},\ell)$ we use the general expression
  \begin{eqnarray}
  \rho_{\rm int}(E^*,J)&=& \frac{1}{16\sqrt{6\pi}}{\left[\frac{\hbar^{2}}{\mathcal{J}_{\parallel}}\right]}^{1/2} a^{-1/4} \nonumber\\
  &\times& \sum_{k=-J}^J [E^*-E_{\rm rot}(k)]^{-5/4}e^{2\{a[E^*-E_{\rm rot}(k)]\}^{1/2}}
  \label{lev_dens}
  \end{eqnarray}
    where is

  \begin{equation}
  E_{\rm rot}(k) =\frac{\hbar^2}{2\mathcal{J}_{\perp}}J(J+1)+\frac{\hbar^2K^2}{2}\left[\frac{1}{\mathcal{J}_{\parallel}} -\frac{1}{\mathcal{J}_{\perp}}  \right];
  \label{ene_rot}
  \end{equation}
  
  in formula (\ref{ene_rot}) $\mathcal{J}_{\perp}$ and $\mathcal{J}_{\parallel}$  are  moments of inertia perpendicular and parallel, respectively, to the symmetry axis, 
and $K$ is the projection of the total spin $J$ on the quantization axis. Application of this general expression depends on the particular cases  \cite{Ignatyuk}. Specific cases take into account: the nucleus at the saddle point,
the case of yrast state, and prolate or oblate or triaxial shape. This expression (\ref{lev_dens}) of $\rho_{\rm int}$ works well for both deformed and spherical nuclei as for example the nuclei very close to the shell closure.

To calculate  $\Gamma_{\rm fis}$ and $\Gamma_{\rm x}$ widths appropriatly, we consider the collective effects in the determination of the level densities through of the non-adiabatic approach (see Appendix B of paper \cite{NPA17}).
In our code the fission and particle decay widths $\Gamma_{\rm fis}$ and $\Gamma_{\rm x}$ are determined by the formulas
\begin{eqnarray}
\Gamma_{\rm fis}(E,J) &=&  \frac{1}{2\pi \rho (E,J)}  \int_0^{E-E_{\rm sad}(J)} \rho_{_{\rm fis}}(E-E_{\rm sad}(J)-\epsilon,J)\nonumber\\
&\times&
 T_{\rm fis}(E-E_{\rm sad}(J)-\epsilon)d\epsilon,
\label{amp_fis}
  \end{eqnarray}
  and
\begin{eqnarray}
\Gamma_{\rm x}(E,J) =  \frac{1}{2\pi \rho (E,J)}\sum_{J'=0}^{\infty}\sum_{j=|J'-J|}^{J'+J} 
 \int_0^{E-B_x} \rho_{_{\rm x}}(E-E_{\rm x}-\epsilon,J')T_{\rm x}^{\ell,j}(\epsilon)d\epsilon,
\label{amp_part}
\end{eqnarray}
where $\rho$ is the level density of a deformed excited nucleus,   $\rho_{_{\rm fis}}$  is the one of the excited nucleus at the saddle point of the fission process, and $\rho_{_{\rm x}}$ is the level density of the successive intermediate excited nucleus after the emission of a particle-$x$ (neutron, proton, $\alpha$-particle, and $\gamma$-quanta).
$E_{\rm sad}$ is the energy of the decaying nucleus at the saddle point with angular momentum $\ell$ and total spin $J$, $T_{\rm fis}$ is the fission transmission coefficient in the Hill-Wheeler approximation and $T_{\rm x}^{\ell,j}$ is the optical-model transmission coefficient for particle-x with angular momentum $\ell$ coupled with particle spin $j$.

Since  for the considered reactions we work in the 10-80 MeV $E^*_{\rm CN}$ excitation energy range, it is necessary to use the non-adiabatic approach for the calculation of the collective level density  $\rho_{\rm coll}^{\rm non-adiab}(E^{*},J)$ (for details see Ref.~\cite{NPA17,NPA1983,Rastop}).

We have to note that in our general model we use the same set of the parameter values for all reactions, and the sensitivity of the final results with respect to the used main parameters have been  presented and discussed in reference ~\cite{NPA17}. In any case, we presented  in Ref. \cite{NPA17} the sensitivity of the functions $P_{\rm CN}$, level density parameter $a$, the $a_{\rm fis}/a_{\rm n}$ ratio, the driving potential and quasifission barrier $B_{\rm qf}$, the capture ($\sigma_{\rm cap}$) and fusion  ($\sigma_{\rm fus}$) cross sections, the damping functions $h(T)$ and $q(\ell)$ to the shell correction in the fission barrier, and the survival probability $W_{\rm sur}$.

\section{Results about the ER excitation functions}

Our codes  allow us to determine  the $\sigma_{\rm ER}$ evaporation residue cross sections  by formula (\ref{equER})
at the given values of the CN excitation energy  $E^{*}_{\rm CN}$ and  at each $x$-step with excitation energy $E^{*}_{x}$  along the deexcitation cascade by emission of light particles (n, p, $\alpha$, and $\gamma$). In order to appropriately calculate the $\Gamma_{\rm x}$ and $\Gamma_{\rm fis}$ widths we consider the collective effects in the determination of the level densities through the use of the non-adiabatic approach
 (see Appendix B of \cite{NPA17}).
\begin{figure}[h!]
\centering
\includegraphics[scale=0.35]{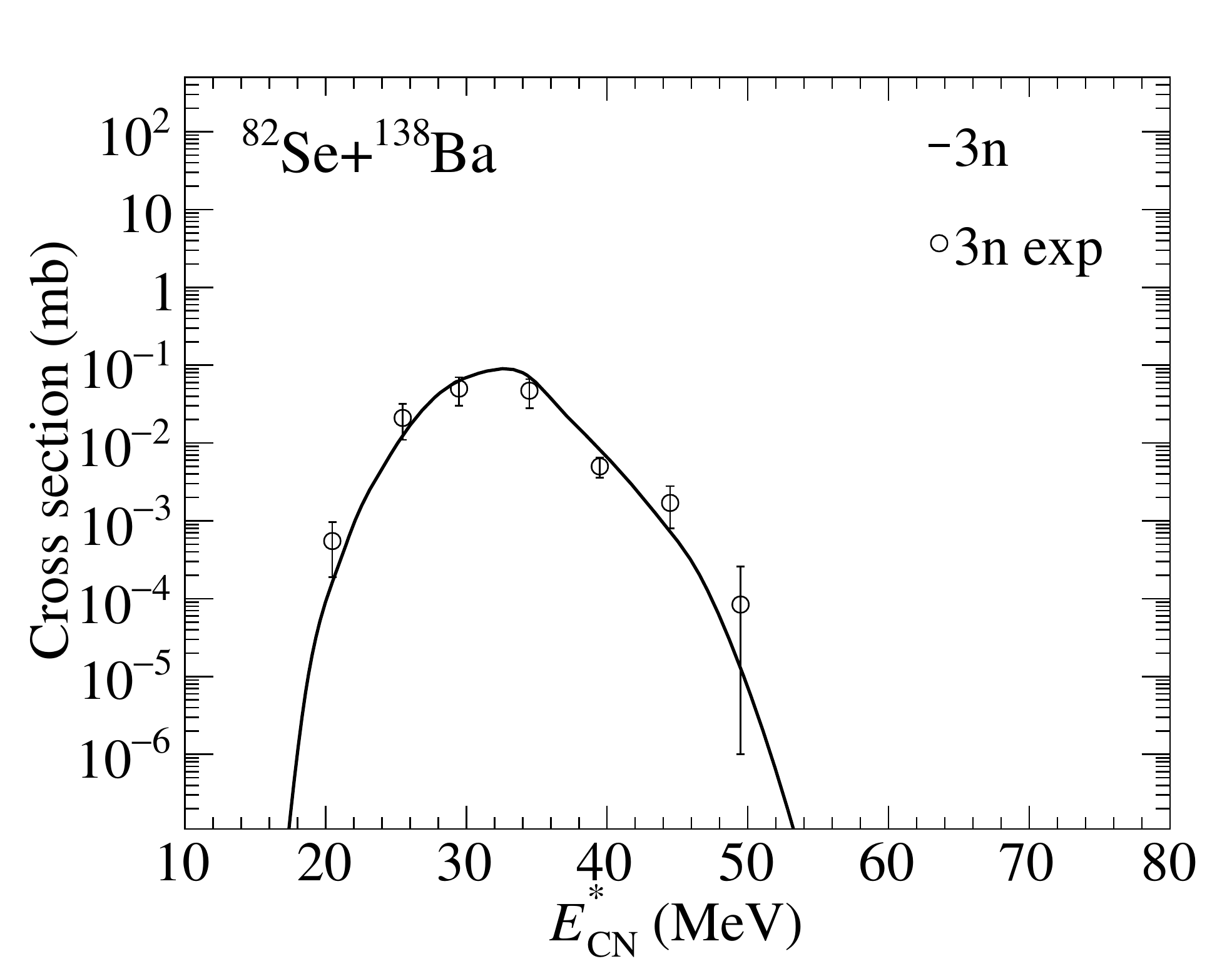}
\put(-40,60){(a)}
\includegraphics[scale=0.35]{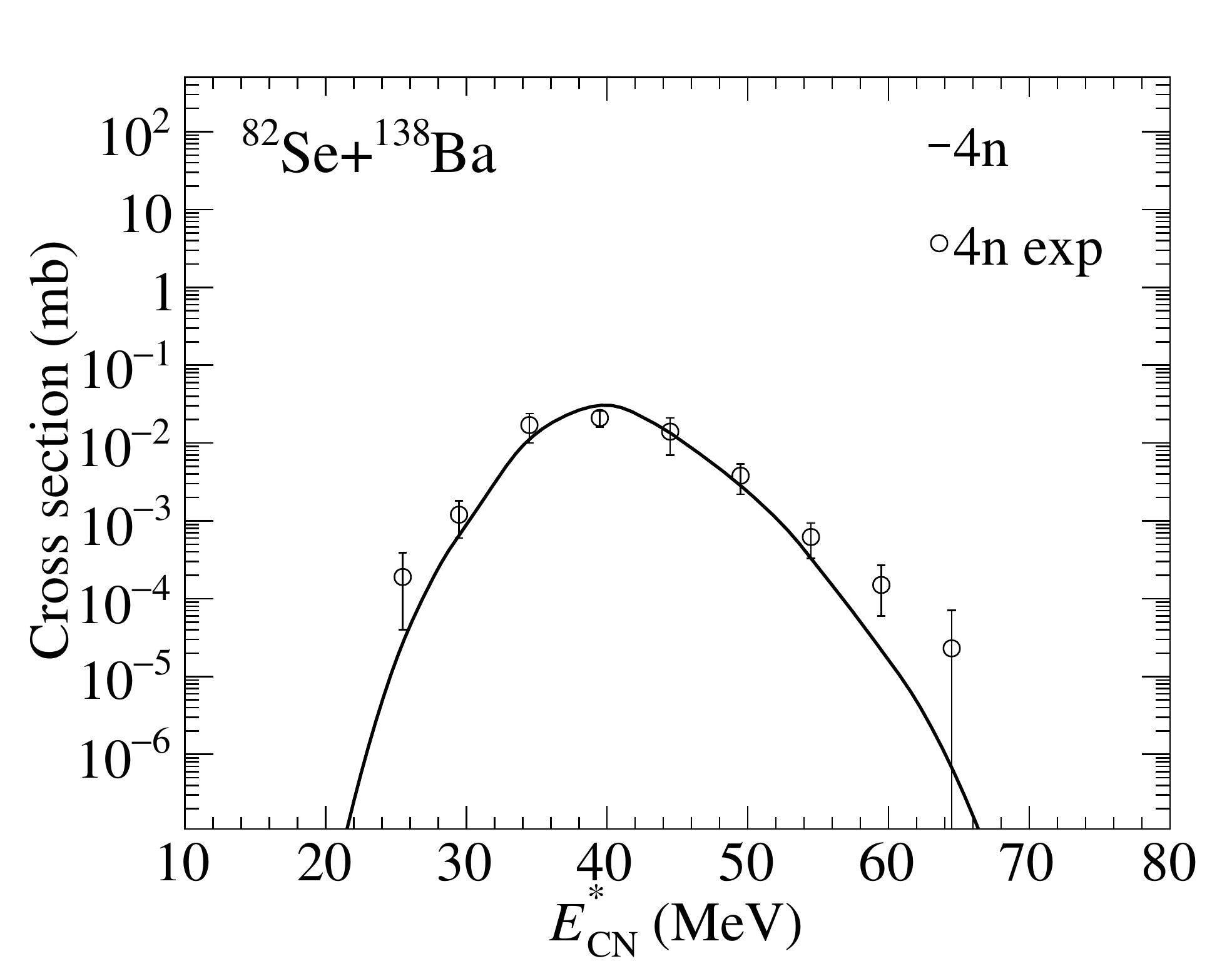}
\put(-40,60){(b)}\\
\includegraphics[scale=0.35]{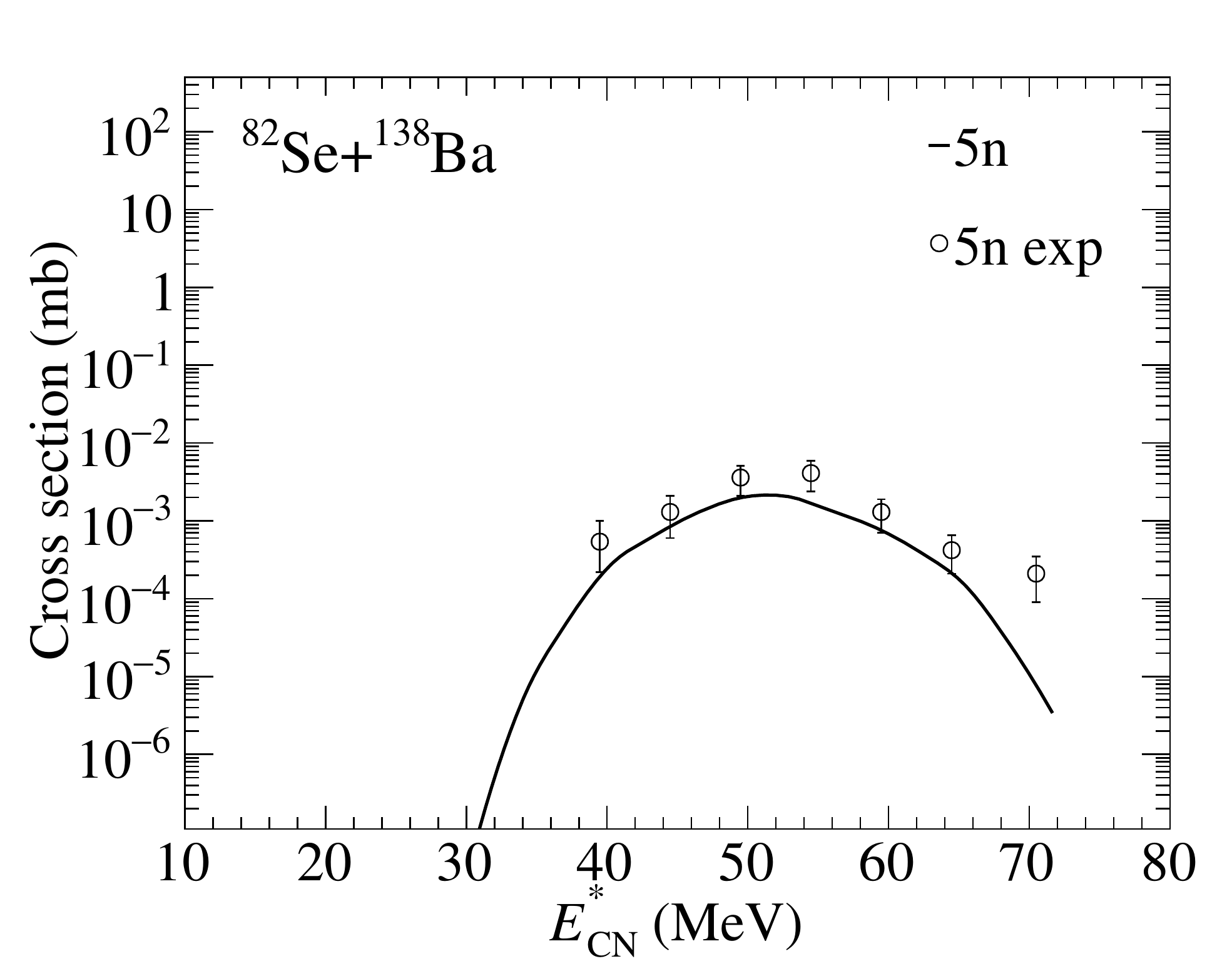}
\put(-140,60){(c)}
\includegraphics[scale=0.35]{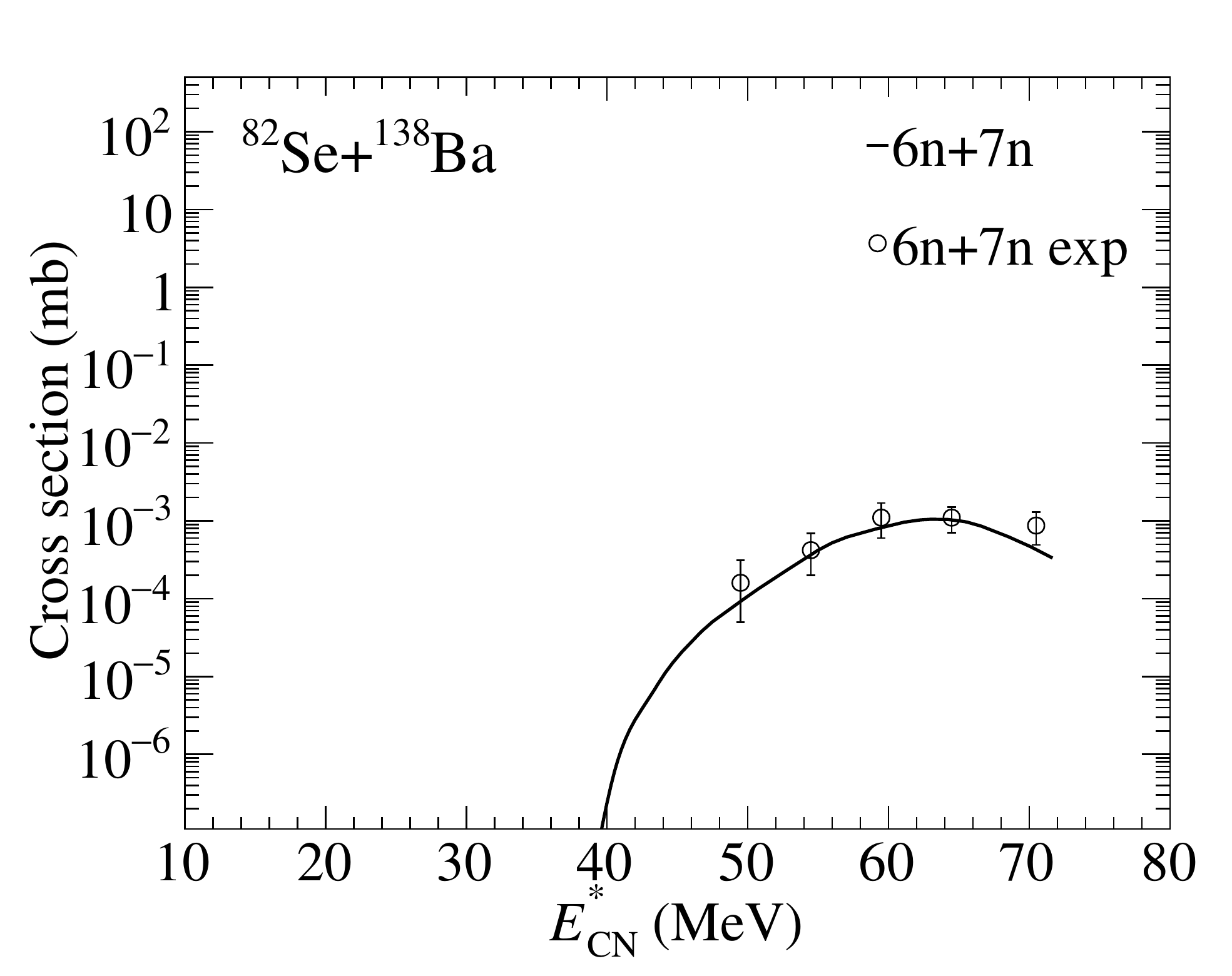}
\put(-140,60){(d)}
\caption{ The individual ER cross sections after 3n (panel (a)), 4n (b), 5n (c),
and 6n+7n (d) neutron emissions vs $E^{*}_{\rm CN}$, the experimental data \cite{PRC652002} are represented by open cirlces while the theoretical estimation by full line.\label{figxn}}
\vspace{-0.3cm}
\end{figure}
In Figure \ref{figxn}, we report the calculated excitation functions of ER cross sections versus $E^{*}_{\rm CN}$ after 3n (panel (a)), 4n (b), 5n (c), and 5n+7n (d) neutron emissions obtained for the $^{82}$Se+$^{138}$Ba reaction leading to the $^{220}$Th CN. The points represents the available experimental values taken form reference \cite{PRC652002}. The results of theoretical calculation are in good agreement with  the available experimental measurements.

\begin{figure}[h!]
\vspace{-0.4cm}
\centering
\includegraphics[scale=0.35]{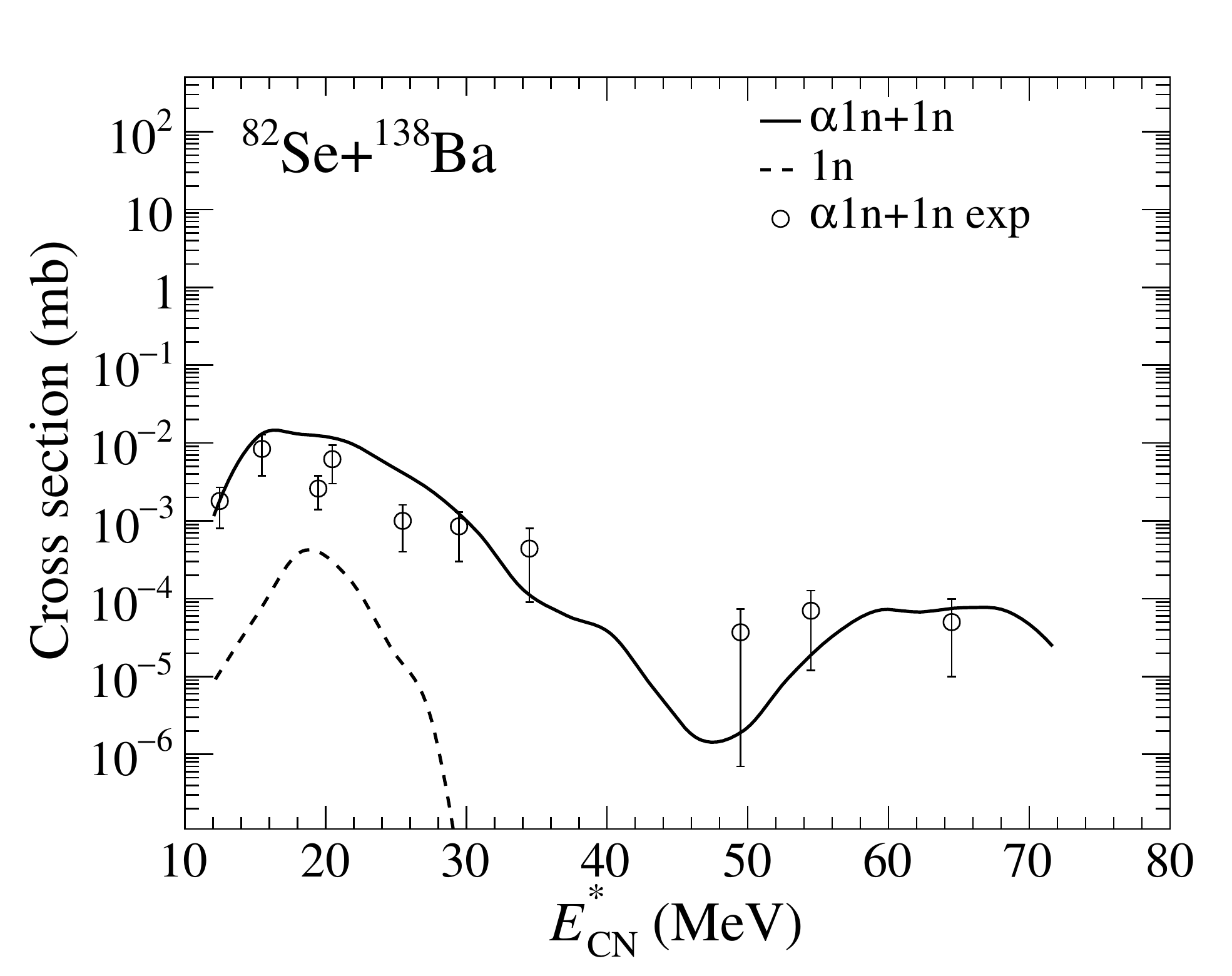}
\put(-40,70){(a)}
\includegraphics[scale=0.35]{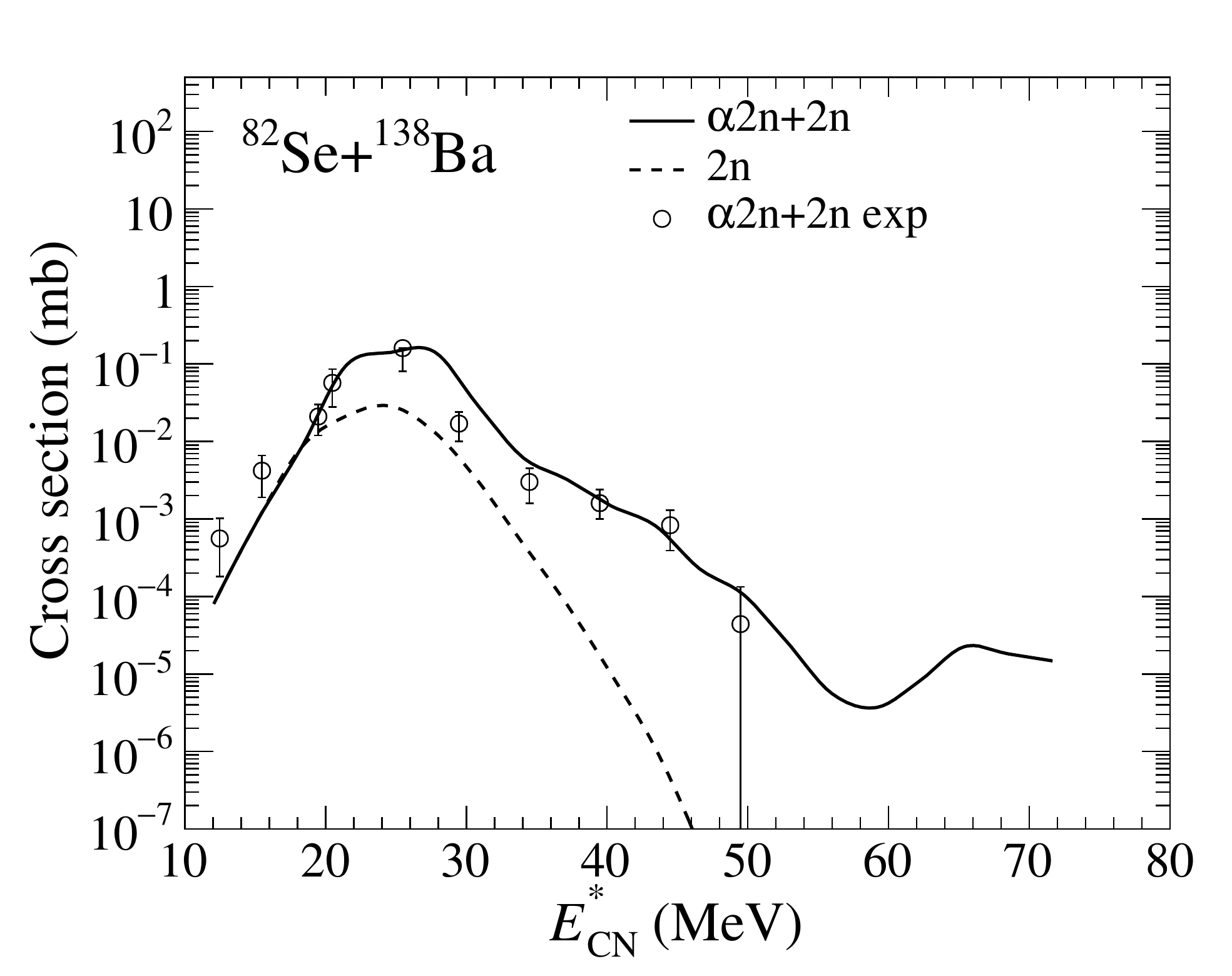}
\put(-40,70){(b)}\\
\includegraphics[scale=0.35]{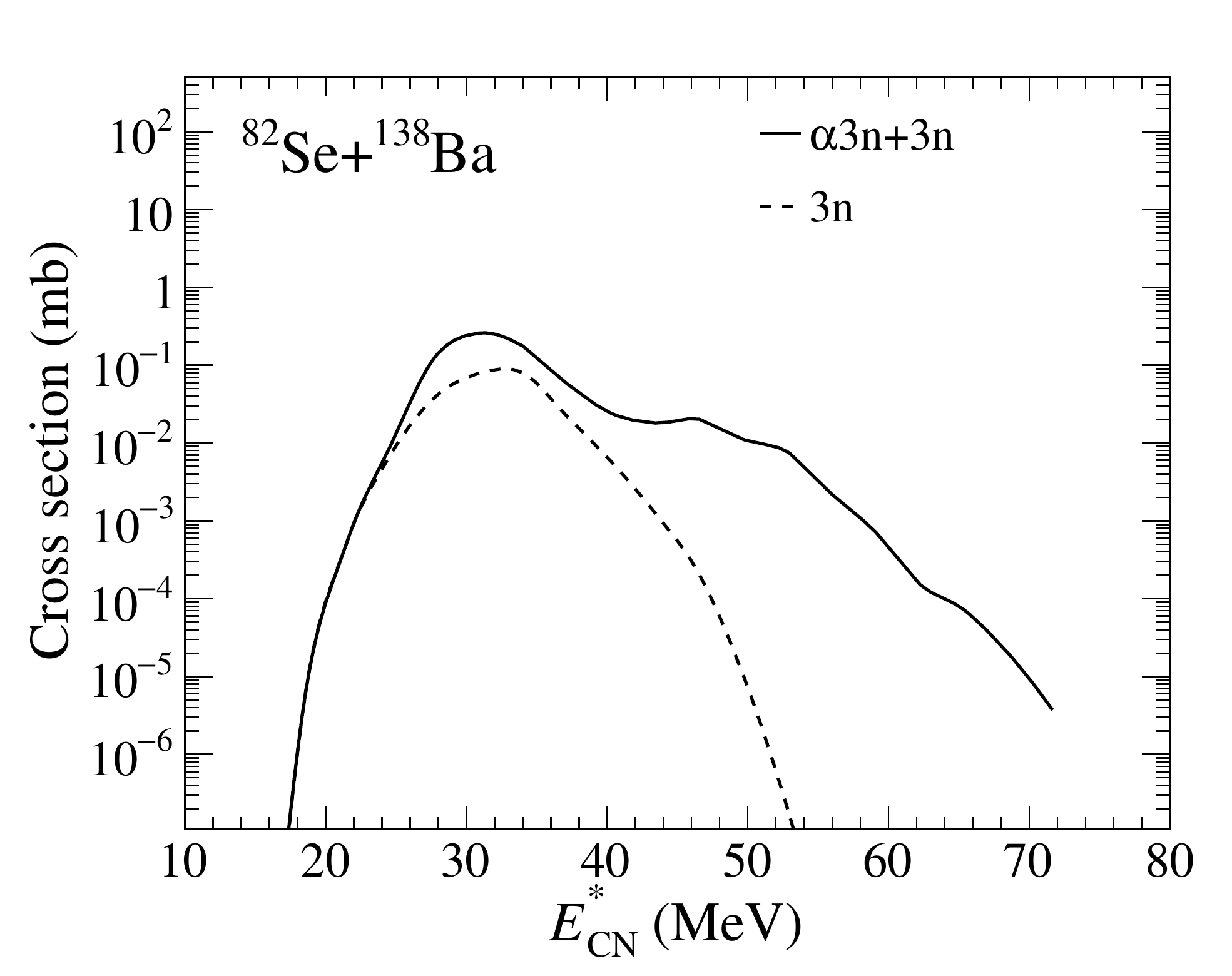}
\put(-40,70){(c)}
\includegraphics[scale=0.35]{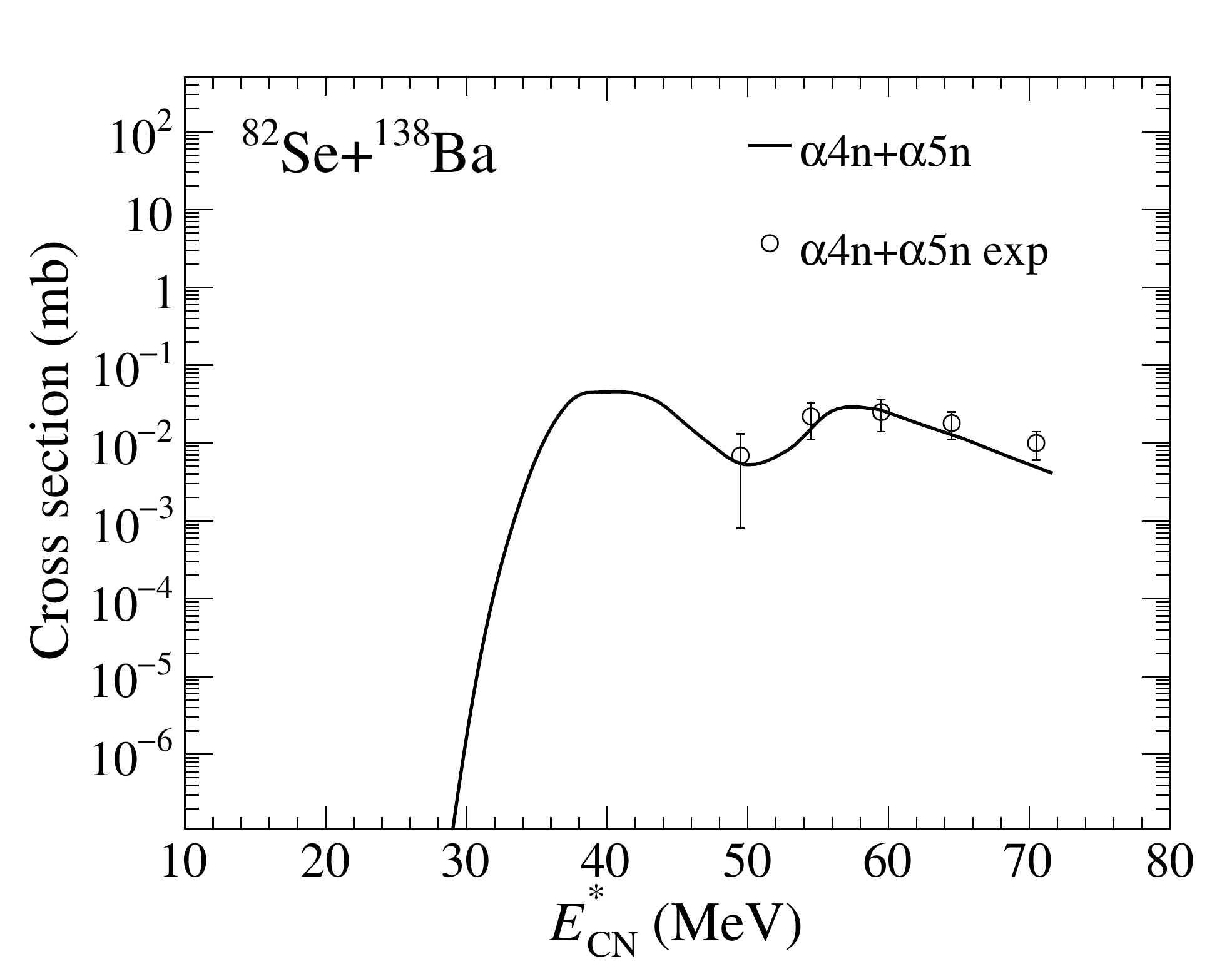}
\put(-150,60){(d)}\\
\includegraphics[scale=0.35]{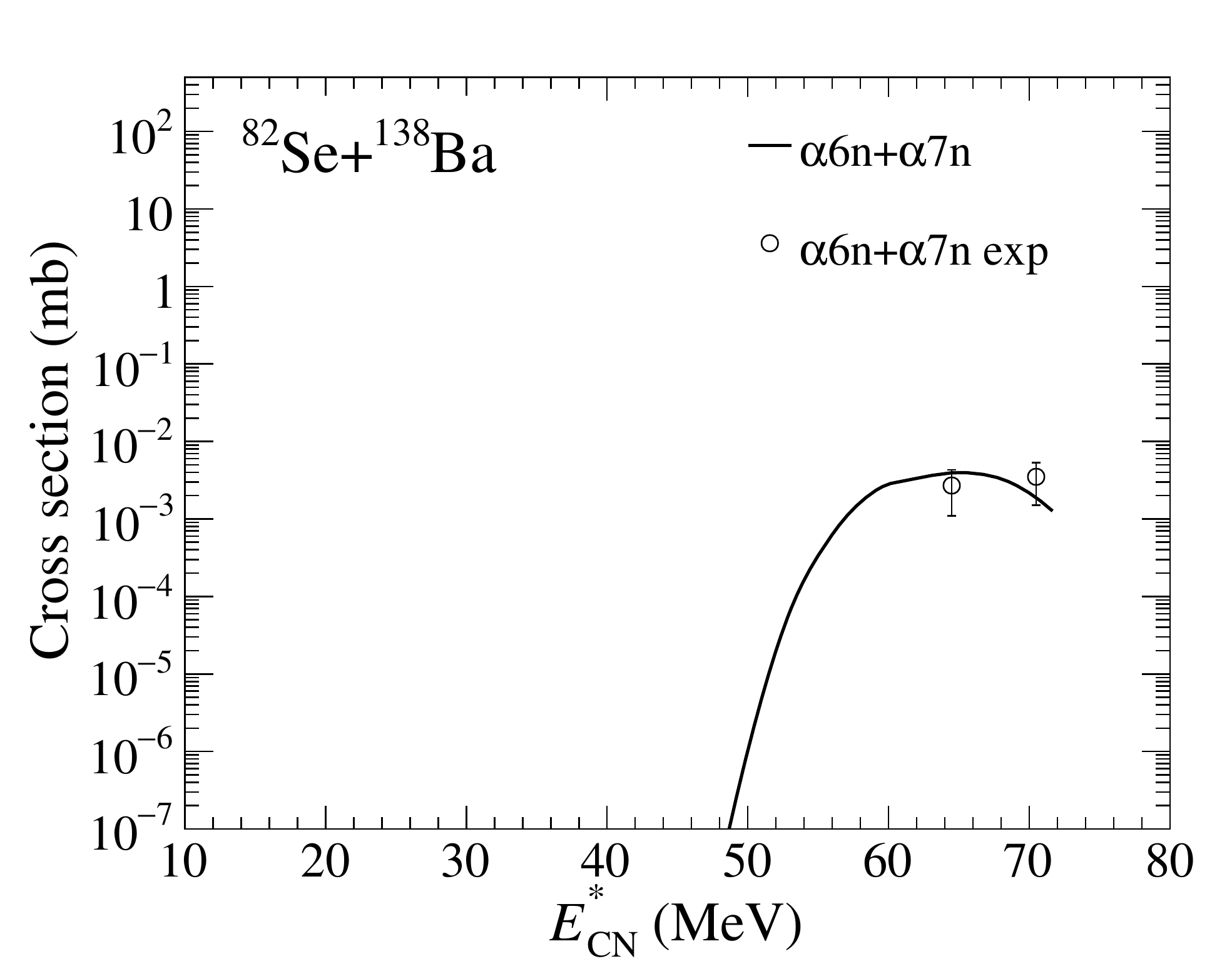}
\put(-150,60){(e)}
\caption{(Color on-line) The individual ER cross sections after $\alpha$1n+1n (panel (a)), $\alpha$2n+2n (b), $\alpha$3n+$\alpha$4n (c),
and $\alpha$6n+$\alpha$7n (d) particle ($\alpha$ and neutrons) emissions vs $E^{*}_{\rm CN}$, the experimental data \cite{PRC652002} are represented by open cirlces while the theoretical estimation by curves (see detail in the insert).\label{figaxn}}
\end{figure}

 Our theoretical analysis  has been extended to  the calculation of   other 
possible relevant
contributions of charged particle emissions as  $\alpha$p$x$n,
2$\alpha x$n,  2$\alpha$p$x$n, 3$\alpha x$n and  3$\alpha$p$x$n leading to the formation
of other various  ER nuclei.
In Figure \ref{figaxn}, we report the calculated excitation functions of ER cross sections
versus $E^{*}_{\rm CN}$ after $\alpha$1n+1n (panel (a)),  $\alpha$2n+2n (b), $\alpha$4n+$\alpha$5n(c) and $\alpha$6n+$\alpha$7n(d) after the indicated light particle emissions. The points represent the available experimental measurements \cite{PRC652002}, and we register the good agreement of the calculated results with the experimental ones.
In panels (a) and (b) we also report by dashed lines the single contributions of 1n and 2n, respectively, in order to highlight the contribution of neutron emission only.

\begin{figure}[h!]
\centering
\includegraphics[scale=0.35]{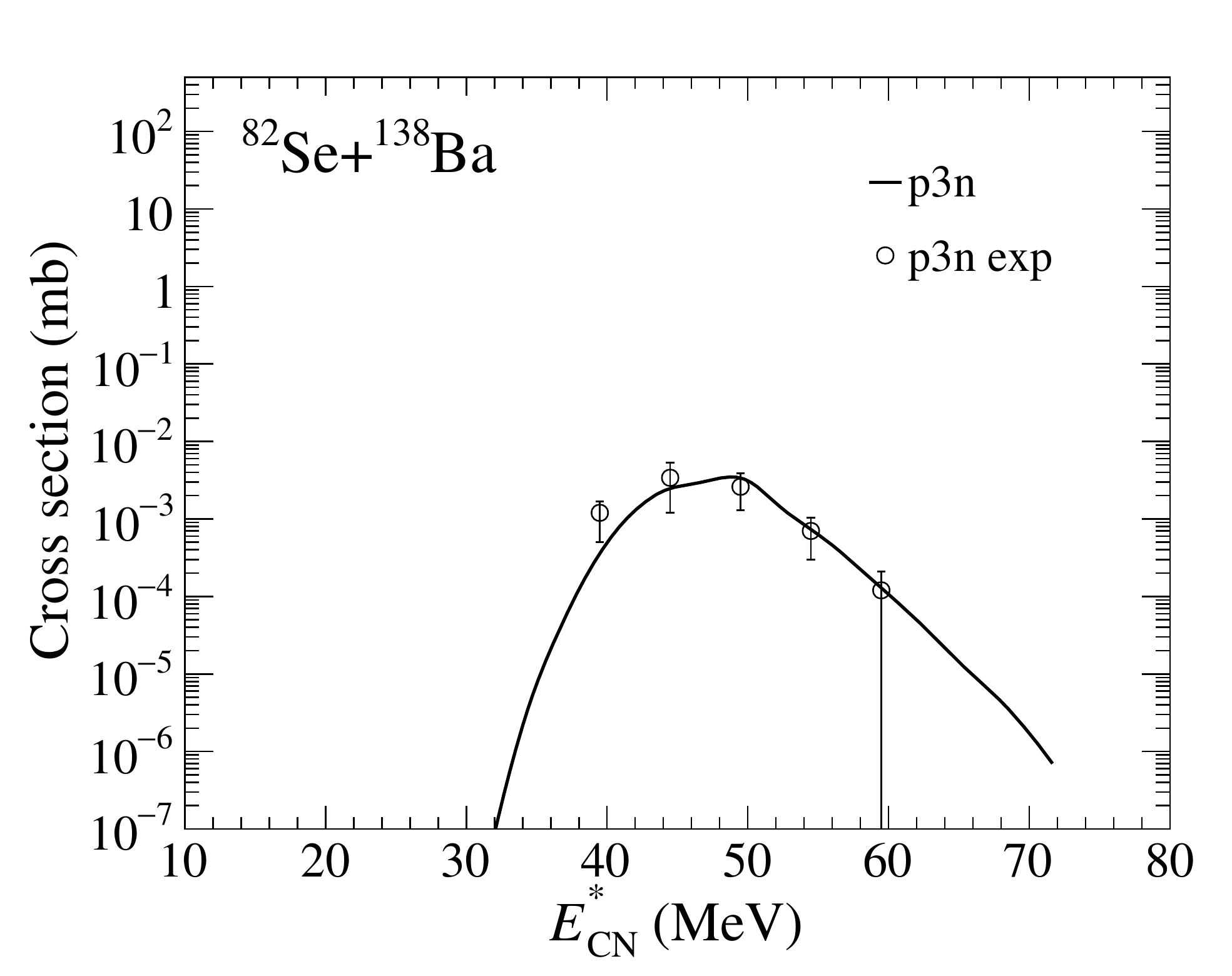}
\put(-160,60){(a)}
\includegraphics[scale=0.35]{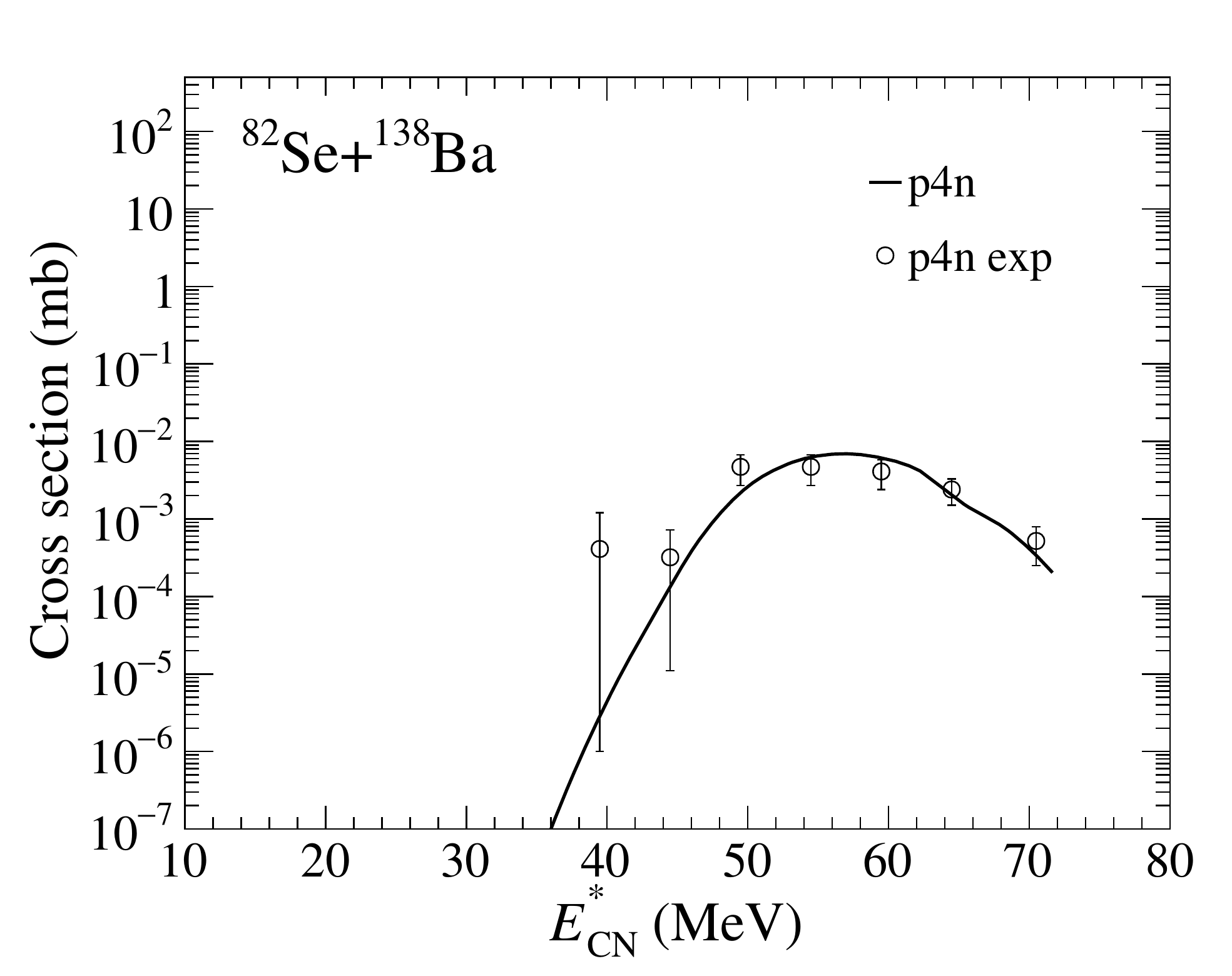}
\put(-160,60){(b)}\\
\includegraphics[scale=0.35]{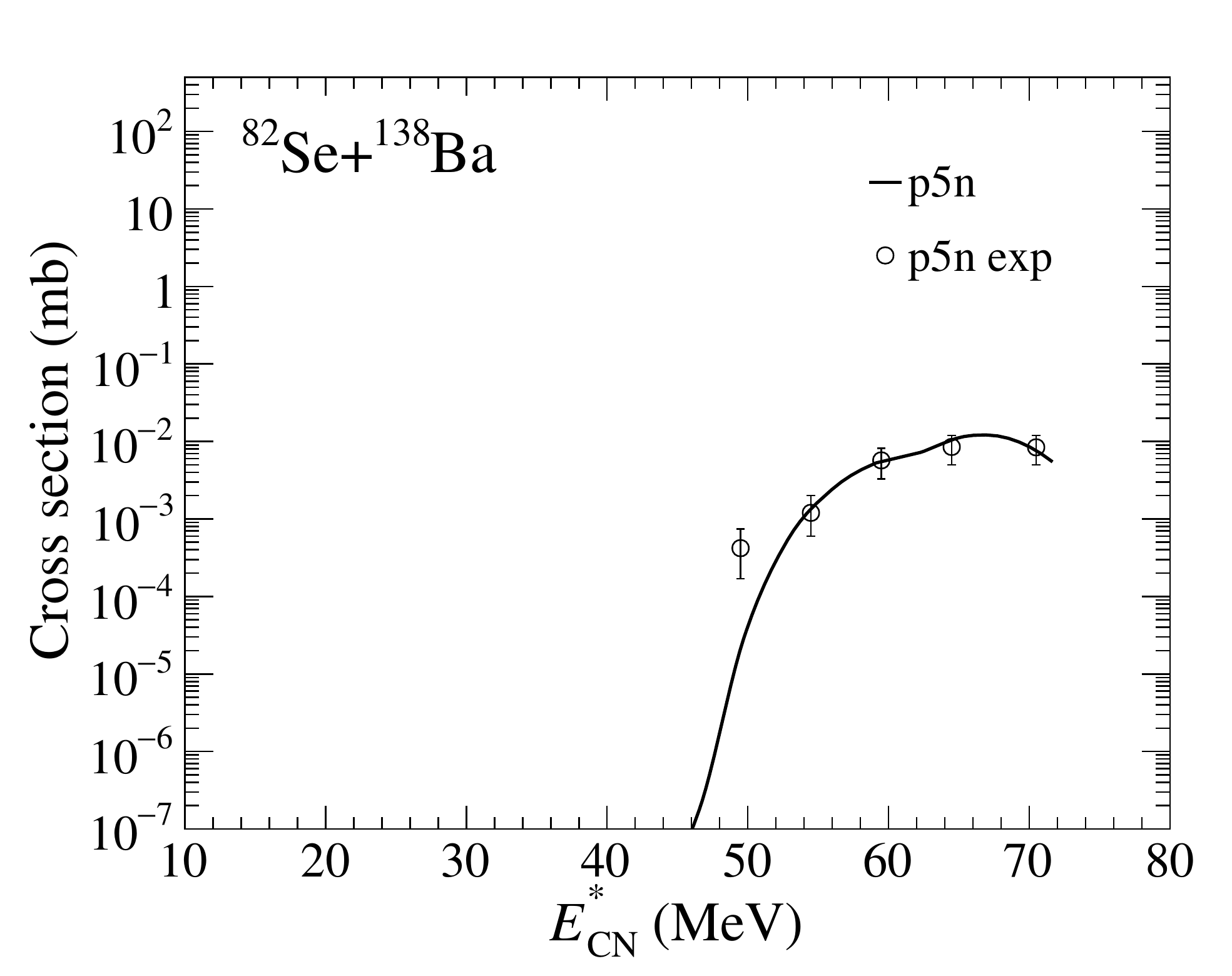}
\put(-160,60){(c)}
\includegraphics[scale=0.35]{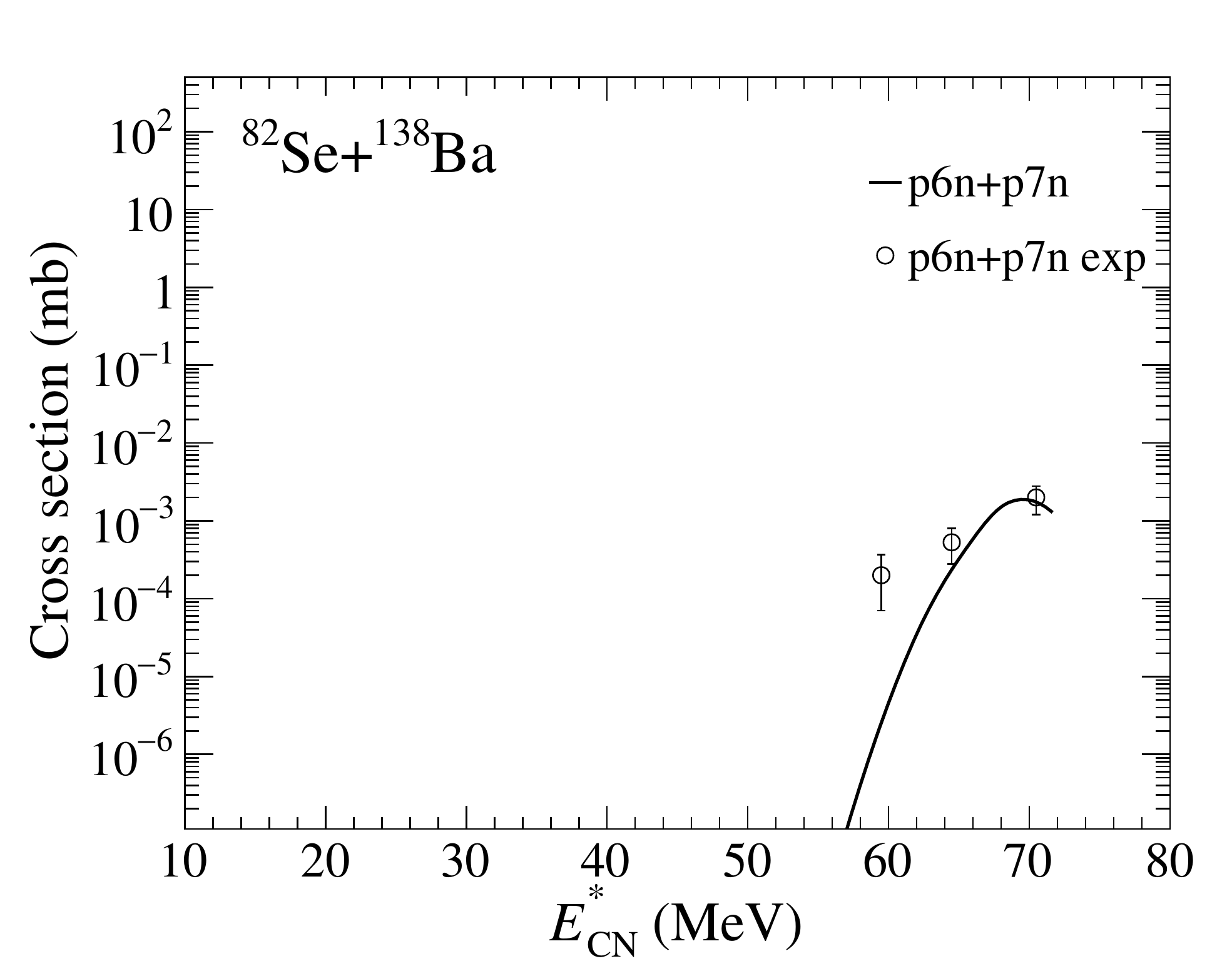}
\put(-160,60){(d)}
\caption{(Color on-line) The individual ER cross sections after p3n (panel (a)), p4n (b), p5n (c),
and p6n+p7n (d) particle (protons and neutrons) emissions vs $E^{*}_{\rm CN}$, the experimental data \cite{PRC652002} are represented by open cirlces while the theoretical estimation by full line.\label{figpxn}}
\end{figure}

In addition, we report in Figure \ref{figpxn} the calculated ER cross sections versus $E^{*}_{\rm CN}$ after p3n  (panel (a)), p4n (b), p5n (c), and p6n+p7n(d)  light particle emissions. Also in this case the points represents the available experimental measurements \cite{PRC652002}, and the corresponding theoretical results are in good agreement with the measurements.
\begin{figure}[h!]
\centering
\includegraphics[scale=0.5]{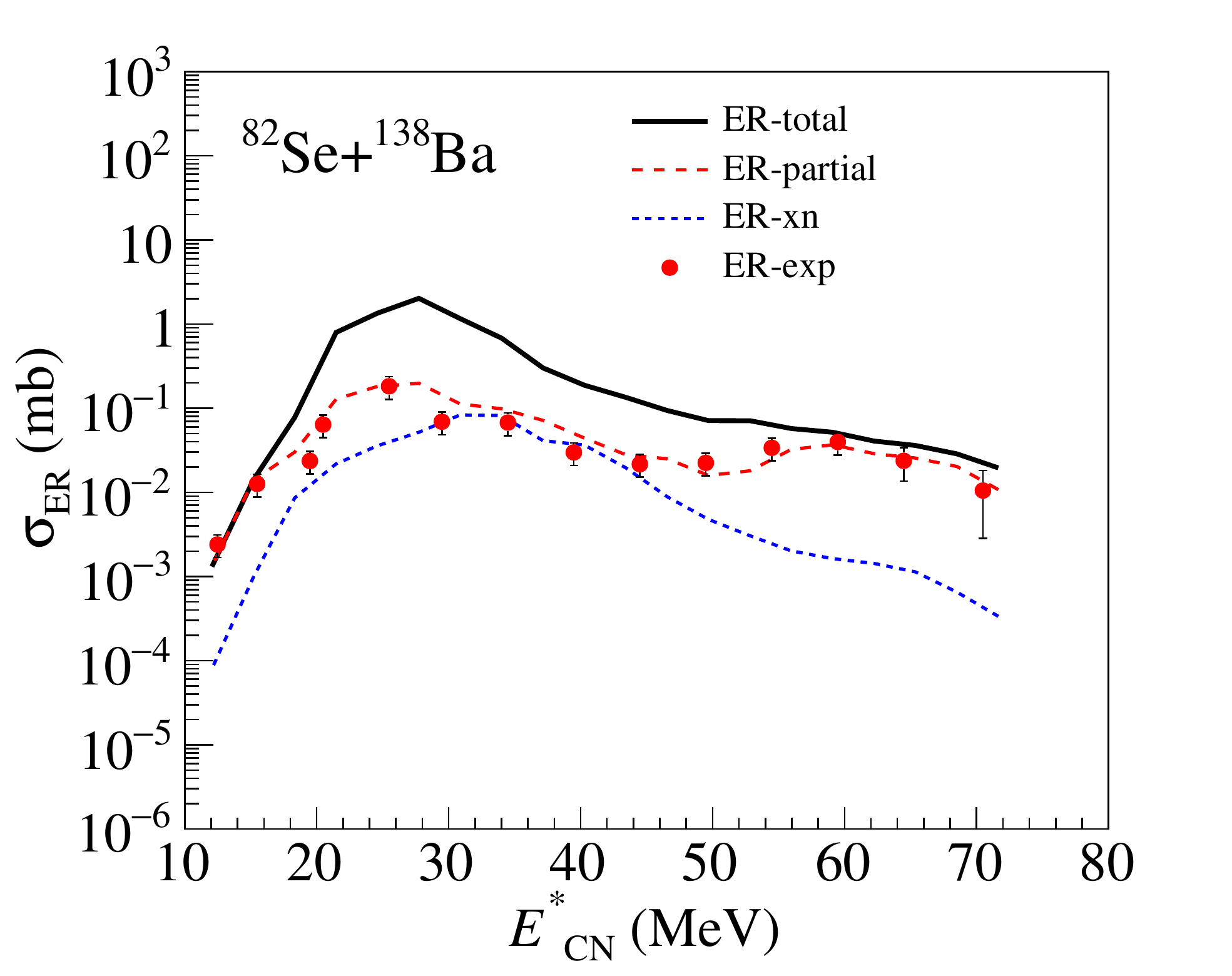}
\caption{(Color online) 
The calculated ER cross sections for the total neutral and charged channels (solid line), the neutron emission only (dotted line),
and the neutral and charged channels corresponding to the measured one  \cite{PRC652002} (dashed line); the experimental ER cross section for the $x$n+$x$pn+$\alpha x$n evaporation channels (full circles) \cite{PRC652002},  for the $^{82}$Se+$^{138}$Ba 
reaction.  
 \label{figER}}
\end{figure}

By observing the complete sets of the experimental results given in Ref.  \cite{PRC652002} for
the  $^{82}$Se+$^{138}$Ba reaction and the corresponding calculated ones we
can affirm that there is a good agreement between the calculated and
experimental results.
However, our theoretical investigation reveals  other important contributions to the total ER cross section, mainly coming from 2$\alpha$xn and $\alpha$pxn channels, according to this we can deduce that the presented experimental results of evaporation residues cover only a partial set of the all possible charged
particle emissions in the investigated $^{82}$Se+$^{138}$Ba reaction. 
In fact, in Figure \ref{figER}, the
  comparison of the theoretical excitation functions of all ER
 relevant decay channels (solid line)
  with the calculated excitation functions of the $x$n+$x$pn+$\alpha x$n evaporation channels only 
  (see dashed line that is in agreement with the experimental data \cite{PRC652002} (full circles))  shows  that the measured excitation function 
  of ER yields formed after the considered charged particle emissions \cite{PRC652002} are lower than 
  our total calculated values in the  17--70~MeV  of  $E^{*}_{\rm CN}$ excitation energy range. We also report in figure the calculated total evaporation residue after neutron emission (dotted line) only.
  
 In fact, the ratios between the values of solid line and the corresponding values of dashed lines are factors 2-3 with peak of 10 at $E^{*}_{\rm CN}=28$~MeV and 4 at $E^{*}_{\rm CN}=50$~MeV, respectively. Instead, the ratios between the values of solid line (the overall ER contributions including the charged emissions too) and the corresponding values of dotted line (representing the ER contributions after neutron emission only) ranges in average within factors 17-53 in the 12-72~MeV interval of $E^{*}_{\rm CN}$  excitation energy.

The discrepancy present in Figure \ref{figER} is due to a wider number of  channels forming the ER contributions taken into account by our theoretical estimations in comparison with those measured experimentally. Our model shows that there are other relevant contributions to the total ER cross section coming from multiple charged emissions like for example 2$\alpha$xn and $\alpha$pxn. This result suggests the importance of improving the experimental possibilities in the identification of each kind of formed ER nuclei, in order to check the reliability of the theoretical models and also to open the way for the research of synthesis of nuclei not directly reachable with the neutron emission only.

\section{Conclusion}

The investigation of the evaporation residue  formation in the 
  almost symmetric  $^{82}$Se+$^{138}$Ba reaction leading to $^{220}$Th CN  has 
  highlighted the role of evaporation residue nuclei obtained after the emission of charged particles on the total ER production. 
  At the same time the recurrent difficulties of detecting and 
  analyzing all the relevant contributions  to the ER formation from the 
  measured  final products of reaction after the emission of the charged
    particles have been discussed.

We have calculated the excitation functions of the ER cross sections after neutron emissions 
only (reported in Figure \ref{figxn}), after $\alpha$xn emissions (reported in Figure \ref{figaxn}), and
 after pxn emissions (reported in Figure \ref{figpxn}). Our estimated ER cross sections are in good agreement with the experimental results given in \cite{PRC652002}, but the comparison  between the calculated excitation function of the total ER production (full line reported  in Figure \ref{figER}) including all contributions of neutron and charged particle emissions and the  measured one \cite{PRC652002}  shows visible differences. In fact, in the 12-70 MeV $E^{*}_{\rm CN}$  energy range the total calculated ER cross section is in average 2-3 times greater than the measured one, while at $E^{*}_{\rm CN}=$28 and 50 MeV the overall calculated ER values are  10 and 4 times greater than the measured cross sections  \cite{PRC652002}. These discrepancies could be due to the experimental limit to measure all relevant contribution to ER cross section formed after many charged particle emissions, in any case the large number of measured ER data, presented in \cite{PRC652002}, offered a good chance to stress about the importance to have also the charged ER data in order to test the reliability of the theoretical model and their ability of predicting the production of nuclei also after the evaporation of charged particles. 

 The presented analysis with the obtained theoretical results for the considered reactions leading to the $^{220}$Th CN can be an useful  information for experimentalists in their investigations on ERs produced by heavy ion reactions, like for example the synthesis of exotic ER nuclei that can not be easily produced by neutron emission only.


\end{document}